\documentclass[preprint2]{emulateapj}
\usepackage{longtable}
\usepackage{cancel}	
\usepackage{color}
\usepackage{graphicx, subfigure}
\usepackage{placeins}
\usepackage{amsmath}
\usepackage{tabulary}

\def\lesssim{\mathrel{\hbox{\rlap{\hbox{\lower4pt\hbox{$\sim$}}}\hbox{$<$}}}}
\def\gtrsim{\mathrel{\hbox{\rlap{\hbox{\lower4pt\hbox{$\sim$}}}\hbox{$>$}}}}

\newcommand{\ips}{\ensuremath{i_{\rm P1}}}

\newcommand{\grizy}{\ensuremath{grizy_{\rm P1}}}

\newcommand{\msun}{\mbox{M$_{\odot}$}}

\newcommand{\kms}{\mbox{$\rm{\,km\,s^{-1}}$}}

\shorttitle{Optical counterpart search for GW151226}
\shortauthors{Smartt et al.}

\begin{document}


\title{A search for an optical counterpart to the gravitational wave event GW151226}


\author{ 
S. J. Smartt\altaffilmark{1}, 
K. C. Chambers\altaffilmark{2},
K. W. Smith\altaffilmark{1},
M. E. Huber\altaffilmark{2},
D. R. Young\altaffilmark{1}
T.-W. Chen\altaffilmark{3},
C. Inserra\altaffilmark{1}, 
D. E. Wright\altaffilmark{1},
M. Coughlin\altaffilmark{4},
L. Denneau\altaffilmark{2},
H. Flewelling\altaffilmark{2},
A. Heinze\altaffilmark{2},
A. Jerkstrand\altaffilmark{1},
E. A. Magnier\altaffilmark{2},
K. Maguire\altaffilmark{1},
B. Mueller\altaffilmark{1},
A. Rest\altaffilmark{5}, 
 A. Sherstyuk\altaffilmark{2},
 B. Stalder\altaffilmark{2},
A. S. B. Schultz\altaffilmark{2},
C. W. Stubbs\altaffilmark{3}
J. Tonry\altaffilmark{2},
C. Waters\altaffilmark{2},
R. J. Wainscoat\altaffilmark{2},
%
%
M. Della Valle\altaffilmark{6,14},
M. Dennefeld\altaffilmark{7},
G. Dimitriadis\altaffilmark{8}, 
R. E. Firth\altaffilmark{8},
M. Fraser\altaffilmark{9},
C. Frohmaier\altaffilmark{8}, 
A. Gal-Yam\altaffilmark{10},
J. Harmanen\altaffilmark{11},
E. Kankare\altaffilmark{1},
R. Kotak\altaffilmark{1},
M. Kromer\altaffilmark{12},
I. Mandel\altaffilmark{13}, 
J. Sollerman\altaffilmark{12},
B. Gibson\altaffilmark{2},
N. Primak\altaffilmark{2},
M. Willman\altaffilmark{2}
}

\altaffiltext{1}{Astrophysics Research Centre, School of Mathematics and Physics, Queens University Belfast, Belfast BT7 1NN, UK}
\altaffiltext{2}{Institute of Astronomy, University of Hawaii, 2680 Woodlawn Drive, Honolulu, Hawaii 96822, USA}
\altaffiltext{3}{Max-Planck-Institut f{\"u}r Extraterrestrische Physik, Giessenbachstra\ss e 1, 85748, Garching, Germany}
\altaffiltext{4}{Department of Physics, Harvard University, Cambridge, MA 02138, USA}
\altaffiltext{5}{Space Telescope Science Institute, 3700 San Martin Drive, Baltimore, MD 21218, USA}
\altaffiltext{6}{INAF, Osservatorio Astronomico di Capodimonte, Salita Moiariello 16, 80131, Napoli, Italy}
\altaffiltext{7}{Institut d'Astrophysique de Paris, CNRS, and Universite Pierre et Marie Curie, 98 bis Boulevard Arago, 75014, Paris, France}
\altaffiltext{8}{School of Physics and Astronomy, University of Southampton,  Southampton, SO17 1BJ, UK}
\altaffiltext{9}{Institute of Astronomy, University of Cambridge, Madingley Road, Cambridge, CB3 0HA, UK}
\altaffiltext{10}{Benoziyo Center for Astrophysics, Weizmann Institute of Science, 76100 Rehovot, Israel}
\altaffiltext{11}{Tuorla Observatory, Department of Physics and Astronomy, University of Turku, V\"ais\"al\"antie 20, FI-21500 Piikki\"o, Finland
}
\altaffiltext{12}{Department of Astronomy and the Oskar Klein Centre, Stockholm University, AlbaNova, SE-106 91 Stockholm, Sweden}
\altaffiltext{13}{School of Physics and Astronomy, University of Birmingham, Birmingham B15 2TT, UK}
\altaffiltext{14}{ICRANET, Piazza della Repubblica 10, 65122, Pescara, Italy}

\begin{abstract}
We present a search for an electromagnetic counterpart of
the gravitational wave source GW151226. 
Using the Pan-STARRS1 telescope we mapped out 290 square degrees in the optical \ips\ filter  starting  11.5\,hr after the LIGO information release  and lasting for a further 28 days. The first observations started 49.5\,hr after the time of the GW151226 detection. 
We typically reached sensitivity limits of \ips$=20.3-20.8$ and covered
26.5\% 
of the  LIGO probability skymap. We supplemented this with ATLAS survey data, reaching 31\% of the probabilty region to 
shallower depths of $m\simeq19$.  
We found 
49
extragalactic transients (that are not obviously AGN), including 
a faint transient in a galaxy at 7\,Mpc  (a luminous blue variable outburst) plus a 
rapidly decaying M-dwarf flare. 
Spectral classification of 
20
other transient events showed them all to be supernovae.
We found an unusual 
transient, PS15dpn,  with an explosion date temporally coincident with GW151226 which evolved into a type Ibn 
supernova.   
The redshift of the transient is secure at 
$z=0.1747\pm0.0001$
and we find it unlikely to be linked, since the luminosity distance has a negligible probability of being consistent with 
that of GW151226.  In the 290 square degrees surveyed we therefore do not find a 
likely counterpart. However we show that our survey strategy would be
sensitive to NS-NS mergers producing kilonovae at $D_{L}\lesssim100$\,Mpc, which is promising for future LIGO/Virgo searches.
 \end{abstract}

\keywords{supernovae: general, supernovae: individual (PS15dpn), 
gravitational waves, surveys}

\section{Introduction}\label{sec:intro}

 The Advanced LIGO experiment detected the first transient gravitational wave signal (GW150914) from the inspiral and merger of 
a pair of black holes of masses 36\msun\ and 29\msun 
\citep{theprizepaper}.  
This was remarkable not only for being the first direct detection of gravitational waves but the first evidence that 
binary black holes (BBH) exist, and the largest mass estimates for black holes in the stellar regime
\citep{2016ApJ...818L..22A}. 
This has been followed by a second discovery, also of a BBH merger signal, with a pair of BHs with masses
$14.2^{+8.3}_{-3.7}$\msun 
and 
$7.5^{+2.3}_{-2.3}$\msun
\citep{gw151226} on 2015 December 26 (GW151226). LIGO estimate a luminosity distance
of
 $440_{-190}^{+180}$\,Mpc
%
corresponding to a redshift  
$z=0.09_{-0.04}^{+0.03}$ 
(90\% limits)
\footnote{Throughout, we adopt the same  cosmological parameters 
as \cite{gw151226} of  
$H_0 = 69\,\kms,\Omega_{\rm M} = 0.31,\Omega_\Lambda =  0.69$. }
%

A broad range of teams  have begun efforts to follow-up GW signals
to detect the putative electromagnetic (EM) counterparts. 
The first event resulted in 25 teams of observers covering the LIGO sky localization 
region with gamma ray to radio facilities 
\citep[summarised in][]{2016arXiv160208492A}.  
The general assumption has been that BBH mergers will not produce a detectable EM signature. 
However Fermi may have detected a weak 
x-ray transient which was temporally coincident with GW150914  
\citep{2016ApJ...826L...6C}, although the reality of the detection 
 is disputed by \cite{2016arXiv160600314G}. 
\cite{2016ApJ...819L..21L} suggested a novel mechanism that may produce both a BBH merger and a relativistic jet from the fragmentation of 
a rapidly rotating core of a single massive star. However if the Fermi hard x-ray detection is real, it is more
like a short gamma-ray burst than a long one.  
Furthermore \cite{2016arXiv160300511W} investigated this scenario quantitatively
and finds a single star origin to be unlikely. 
\cite{2016ApJ...821L..18P}  proposed a short GRB may be formed
if the two black holes are formed within a 
fossil disk which restarts accretion due to  tidal forces and shocks during the BBH merger.
Hence the searches continue, particularly as a detection would open up a major new way to probe high energy astrophysics, 
stellar evolution, compact remnants and test modified theories of gravity \citep{2016JCAP...03..031L}. 

Here we present the results of our wide-field search for an optical counterpart to the 
transient gravitational wave event GW151226 using the Pan-STARRS1 (PS1) and the ATLAS survey telescopes combined with spectroscopic
follow-up from Hawaiian facilities and the Public ESO Spectroscopic Survey of Transient Objects (PESSTO).

\begin{figure*}
\begin{center}
\includegraphics[width=11cm,angle=0]{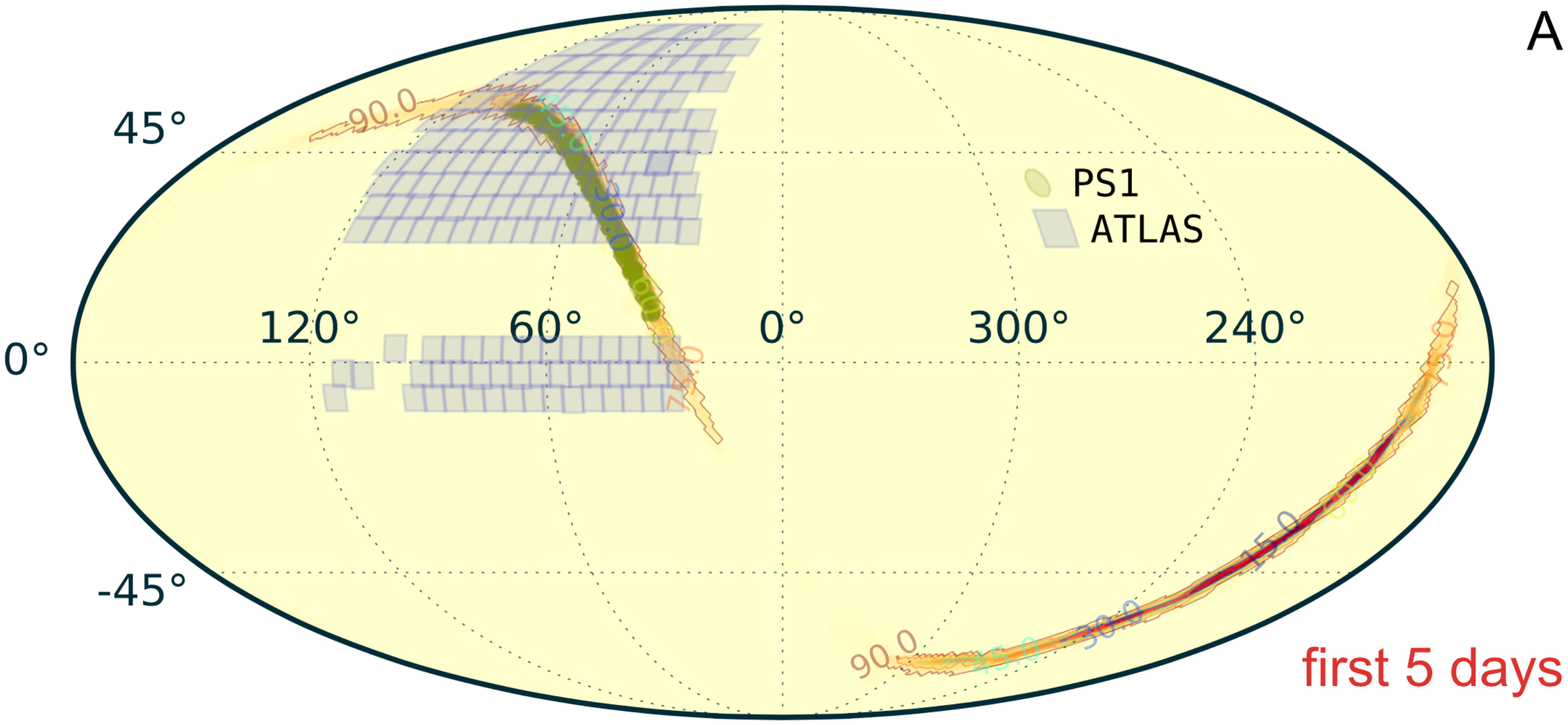}
\end{center}
\includegraphics[width=7.5cm,angle=0]{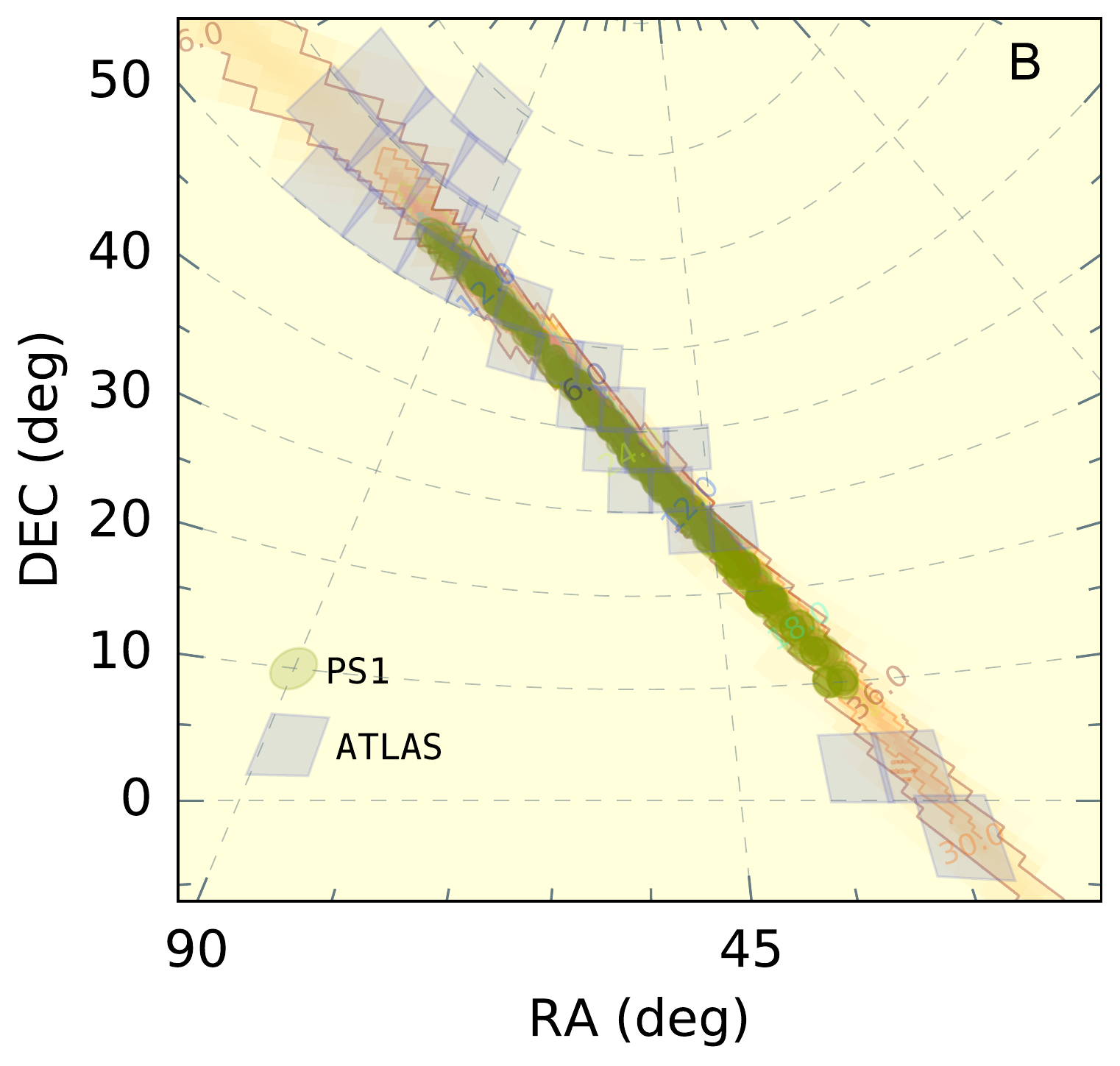}
\includegraphics[width=10.5cm,angle=0]{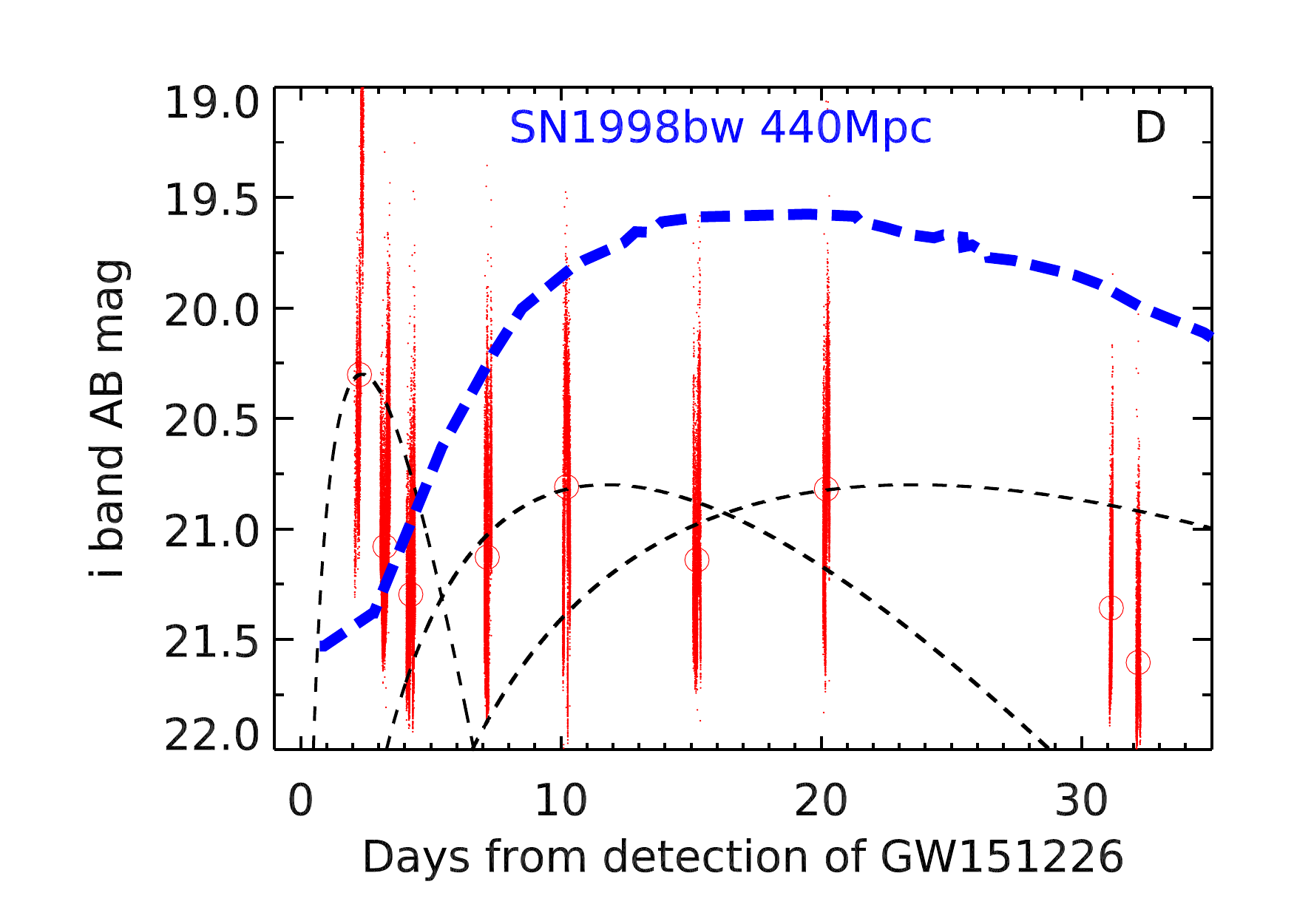}
\includegraphics[width=7.5cm,angle=0]{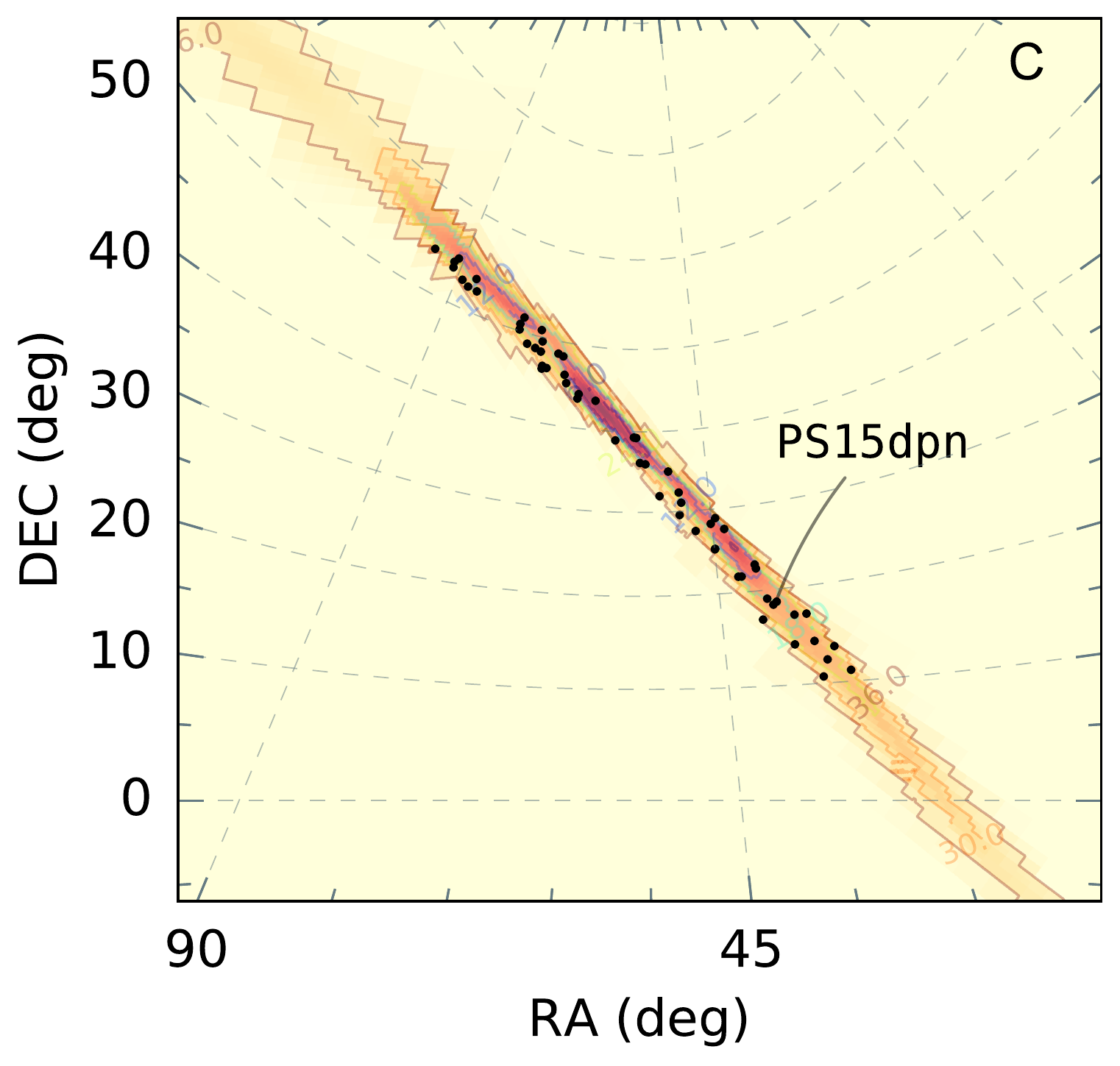}
\includegraphics[width=10.5cm,angle=0]{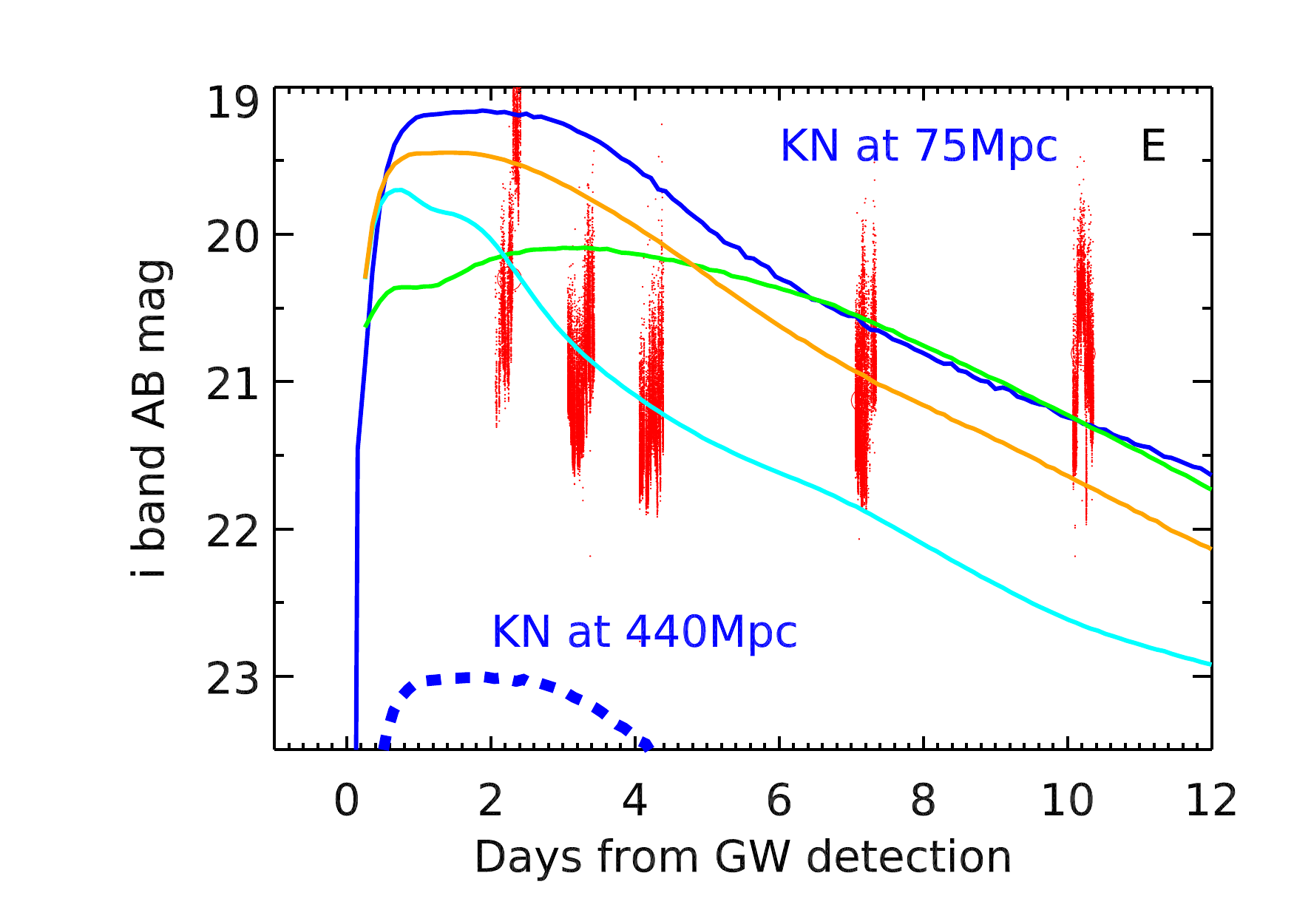}
\caption{{\bf A:} LIGO sky localization region showing Pan-STARRS1 and ATLAS sky coverage within 5 days of GW151226. 
{\bf B and C:} Zoom in of our focused region with Pan-STARRS1 and ATLAS sky coverage with transients detected. 
{\bf D:} 5$\sigma$ detection limits for all \ips images,  with the median nightly value marked as red open circles. The 
black curves are parameterised lightcurves of three different timescales (4d, 20d, 40d) and the blue line is SN1998bw 
placed at $D_{\rm L}=440$\,Mpc. 
{\bf E:} NS-NS mergers expected to be detected within $D_{\rm L}=75$\,Mpc, as expected in the upcoming 2016 LIGO-VIRGO run. 
Our 5$\sigma$ limits are shown with kilonova models.  Blue: the disc wind outflows of compact object mergers of
\cite{2015MNRAS.450.1777K}.  Red: r-process powered merger model which includes a $^{56}$Ni-dominated wind 
\citep{2013ApJ...775...18B}. Cyan:  merger model with iron-group opacity with 
$M_{\rm ej} =0.01$\msun\ by the same authors. Green: merger model for opacity dominated by r-process elements, with $M_{\rm ej} =0.1$\msun\  
also by the same authors. All in SDSS-like $i-$band, AB mags. }
\label{fig:skymap}
\end{figure*}

\section{Observing campaign of  source GW151226}\label{sec:obs}

To search for optical counterparts to gravitational wave events our collaboration 
\citep{2016arXiv160204156S} uses the Pan-STARRS1 system  \citep{2010SPIE.7733E..12K}
for imaging and  relies on the existence of the Pan-STARRS1 3$\pi$ Survey 
(Chambers et. al. 2016 in prep) for template images. 
The PESSTO Survey \citep{2015A&A...579A..40S} 
together with programs on Gemini North with GMOS, the UH2.2m with SNIFS,  
provide spectroscopic classification. The data for one object discovered here were
supplemented  with Hubble Space Telescope observations. 
GW151226 was detected on 
2015 December 26 03:39 UTC (MJD 57382.152)
and released to the EM community as a discovery on
2015 December 27 17:40 UTC \citep{gw151226}. 
The initial localization generated by the BAYESTAR pipeline 
\citep{2016PhRvD..93b4013S}
contained a 50\% credible region of 
430 square degrees and a 90\% region of about 1400 square degrees 
\citep[to be compared with 90\% credible region of 630 square degrees for GW150914][]{2016arXiv160208492A}.
We began taking data with the Pan-STARRS1 telescope during the next available dark hours, 
on  2015 Dec 28 05:08 UTC 
(11.47\,hr after the LIGO information release and 49.48\,hr after the event time)
and mapped out a region of 214 square degrees
on this first night as shown in Figure\,\ref{fig:skymap}. 

The same region was mapped on the two subsequent nights (extending to 273 square degrees). All observations were done with the 
Pan-STARRS1 \ips\ filter with a 4-point
dither pattern at each pointing centre. The four individual (back to back) 45 sec exposures 
were co-added to produce a 180s exposure and the Pan-STARRS1 3$\pi$ 
\ips reference image  (typically having an effective total exposure time of 270-900 sec) 
was subtracted from this 180s night stack \citep[see][and Magnier et al. in prep, for more details]{2016arXiv160204156S}. 
On any one night this 
180s exposure sequence was repeated
 multiple times (2-3) in the central highest probability region, giving us some intra-night 
sampling. 
The sequence was repeated a further 5 times between 2016 January 02 and January 25 (extending the full footprint to a total of 290 square degrees). The observing cadence and sensitivity  are illustrated in Figure\,\ref{fig:skymap}, and the full 
PS1 footprint corresponds to
26.5\% of the full LIGO posterior probability. 
This footprint choice was a
combination of telescope accessibility of the LIGO localization map  and a choice to go deeper on the
higher probability regions \citep{2016arXiv160405205C}. 

We selected targets with similar filtering algorithms as described in our first paper \citep{2016arXiv160204156S}. 
A total of 2.3$\times10^{7}$
detections were ingested into the database  (after basic rejections of known defects).
Spatial aggregation of detections within 0\farcs5  of each other resulted in the creation of 
1.1$\times10^{7}$ objects 
and basic filtering and insistence of two separate detections resulted in a total of 
1.7$\times10^{6}$
candidate astrophysical transients. 
Subsequent 
filtering (obvious dipoles, stellar objects and objects near bright stars) and a random forest machine learning classifier reduced the numbers to 
144,000
for which the pixel recognition machine learning technique was employed \citep{2015MNRAS.449..451W,2016arXiv160204156S}. 
Further removal of 
3,903
known minor planets left a total of 
24,100
objects for humans to scan and this manual process resulted in 
85
objects for further investigation.   
The human scanning involved removing artefacts that are obvious to the 
eye but are not properly recognised by the machine learning. As we wanted to 
err on the side of completion over purity, we set the machine learning 
threshold to roughly a 20 per cent false positive rate on the ROC curve
\citep[see Fig.7 of][for an illustration]{2015MNRAS.449..451W}. 
The human scanning removed subtraction and chip defects that are
easily distinguished visually. 
We note that in the Milky Way plane
there were at least a further 43 faint transients which are very likely variable stars that 
reach above our detection limit on a few epochs.  A few could be 
background hostless supernovae, but their location in the plane suggests
they are faint stellar variables. 

In addition, the ATLAS 0.5m telescope \citep{2011PASP..123...58T}, covered a significant fraction of the northern sky  in the first five days after GW151226 as shown in Figure\,\ref{fig:skymap}. These data 
were taken during normal ATLAS operations and can be thought of as ATLAS working in 
serendipitous  mode. In future, ATLAS will be able to work in targeted mode in the same 
way as Pan-STARRS1. A single ATLAS unit, with its 30 square degree cameras can map out 1000 square
degrees within 30 minutes. 
We highlight that  just 3\,hrs after the GW151226 event detection,  ATLAS  serendipitously covered 87 square degrees of the 
sky localization region (2.2\% enclosed probability) 
during the time window $57382.302\pm0.014$. 
We processed all ATLAS 
data taken serendipitously in the first 5 days to locate transients as in \cite{2016ATel.8680....1T}. After processing about 575 sq. degrees, the ATLAS coverage increases the total enclosed probability to 36\% 
over the first 5 days from GW151226, getting to median 5$\sigma$ limits of $m_{o}\simeq19.0$ (orange filter). Apart from variable stars
and CV candidates, we found no other extragalactic transient candidates in this stream.

\begin{figure}
\includegraphics[width=\columnwidth,angle=0]{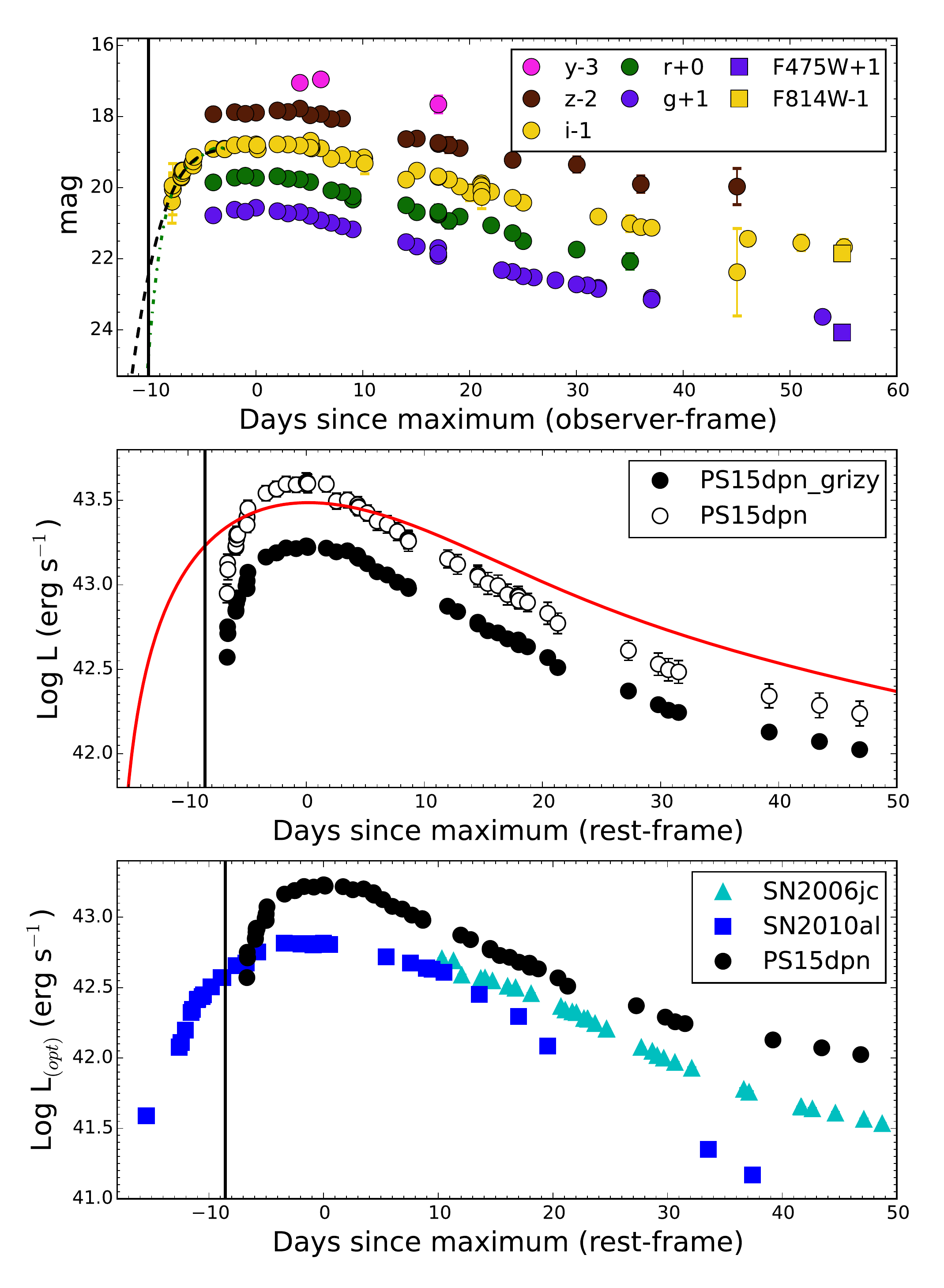}
\caption{{\bf Upper:} PS1 lightcurve of PS15dpn. Circles are PS1 and square symbols are from HST. 
Black dashed line is 3rd order fit and green dot-dashed is 4th order (each fit to first  6 epochs). Vertical black 
line indicates time of GW151226 event. 
{\bf Middle:} Bolometric luminosity calculated with \grizy filters only \citep[see][for details]{2016arXiv160401226I}
and a full bolometric lightcurve estimated from a black body extrapolation
between $0.2-2.5\mu$m. A simple Arnett model, as described in \cite{2013ApJ...770..128I} is shown for the latter
with input parameters: $E_{\rm exp}=5\times10^{51}$\,ergs, $M_{\rm ej}=1.9$\msun, $M_{Ni}=1.7$\msun. 
This simply indicates that a $^{56}$Ni model is a 
poor fit and that type Ibn are not well explained by
radioactive powering. 
{\bf Lower:} comparison with two  well observed SN Ibn \citep{2007Natur.447..829P,2015MNRAS.449.1921P}.
}
\label{fig:15dpn-lc}
\end{figure}

\subsection{Discovery and spectroscopic classification of transients}\label{sec:class}

During our filtering, we removed obvious Galactic stellar variables and known AGN candidates. 
After removing these contaminants, we found 
49 transients which are either confirmed SNe or likely SNe which are all summarised in 
Table\,\ref{tab:trans}.  As discussed in \cite{2016arXiv160204156S} 
the detected transients are dominated by
mostly old supernovae that exploded over an extended period before the GW trigger. 
The sky position of transients found in the first three days are  plotted in Figure\,\ref{fig:skymap}. 
Those with spectroscopic classifications are listed along with their redshifts.  We suggest that all these objects are unrelated field supernovae, although  one object, PS15dpn deserves closer inspection and is discussed in the next section.

We note two objects that are unrelated to GW151226 but are worth  highlighting in the context of searching for unusual transients in  LIGO/Virgo sky localization regions.
PS15dqa is a faint transient  in the nearby ($D=7$\,Mpc) galaxy NGC 1156. The transient magnitude  \ips=20.8  implies $M_i = -8.9$ (including significant Milky Way foreground extinction of $A_i = 0.36$). A Hubble Space Telescope (HST) archive image with the Advanced Camera for Surveys (ACS, in filter F625W) of NGC~1156 shows an object which is astrometrically coincident with PS15dqa to within 0\farcs3. This is a 
stellar point source with $m_{\rm F625W} = 20.1$ and hence $M_{r} =-9.6$. Assuming a bolometric correction of 
zero this corresponds to $\log L/{\rm L_{\odot}} = 5.7$\,dex which implies a 50-60\msun\ star that is 
undergoing brightness variations by a factor of 2. While this scale of variability is known 
for massive stars of this luminosity, and is unrelated to GW151226, it illustrates our ability to identify faint transients in nearby galaxies. 
Secondly,  PS16li is a fast optical transient with a 1.3 magnitude fade in 13min on the night of  MJD=57397.51 with a 
faint, red point source in the PS1 reference stack. 
This is an M-dwarf flare \citep[e.g. ][]{2013ApJ...779...18B} 
which highlights our ability to pick up 
fast decaying transients.

\begin{figure}
\includegraphics[width=\columnwidth,angle=0]{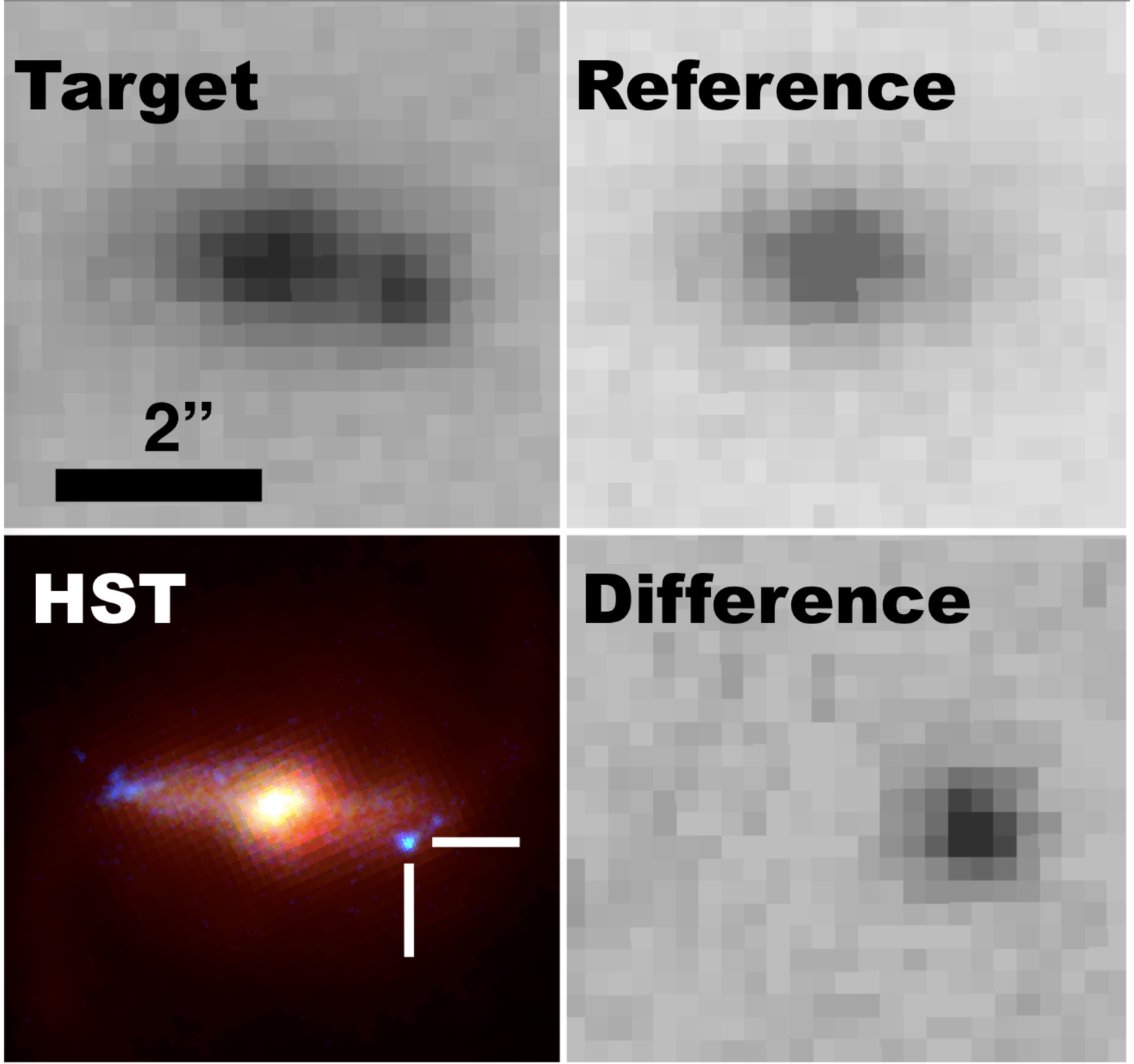}
\caption{Target, reference and difference \ips images from PS1 on MJD=57386 showing PS15dpn offset from its host galaxy. Color composite (F475W, F814W, F160W) image from HST on MJD=57447.}
\label{fig:15dpn-hst}
\end{figure}

\begin{figure*}
\includegraphics[width=16cm,angle=0]{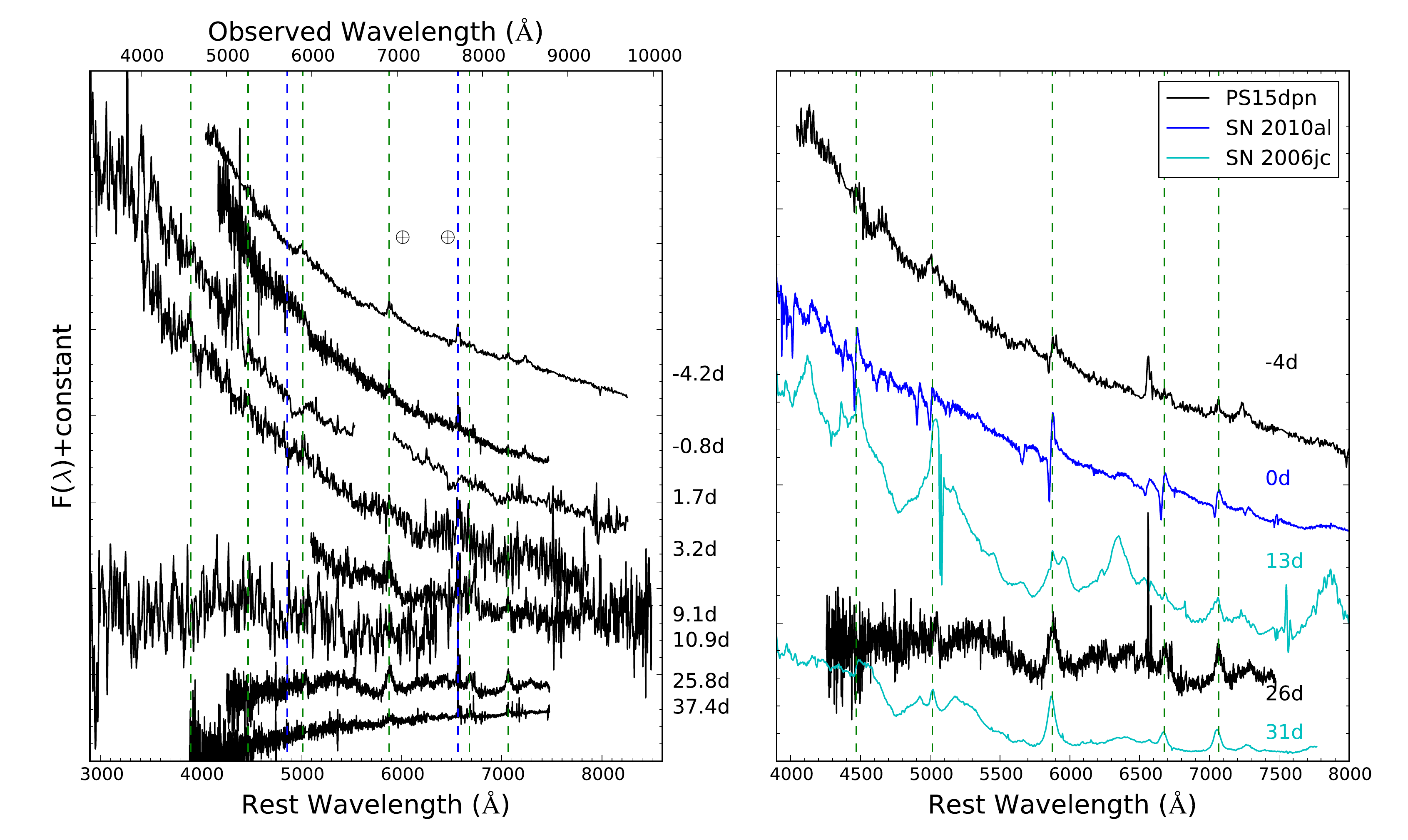}
\caption{Spectra of PS15dpn from the combined GMOS, PESSTO and SNIFS campaign. 
The vertical dashed green lines refer to He\,{\sc I} and He\,{\sc II} lines, while the blue (only shown on left) refer to H$\alpha$ and H$\beta$. Right panel refers to restframe days after peak.}
\label{fig:15dpn-spec}
\end{figure*}

\subsection{PS15dpn : a type Ibn supernova temporally coincident with GW151226}
\label{sec:ps15dpn}

This object was reported early in the campaign as being of interest because of its rising light curve and 
very blue spectrum \citep{GCN18811}.
We gathered a multi-color PS1 lightcurve in \grizy\ 
\citep{2012ApJ...750...99T,2013ApJS..205...20M,2012ApJ...756..158S}
and one epoch of HST imaging with WFC3 (see Figures\,\ref{fig:15dpn-lc} and \ref{fig:15dpn-hst}), together
with 8 epochs of spectra.  
The redshift of the host galaxy
is  measured at $z=0.1747\pm0.0001$ ($D_L = 854$\,Mpc) from the centroids of 
the strong host galaxy emission lines of H$\alpha$, [N\,{\sc ii}] and [S\,{\sc ii}]. 
Figure\,\ref{fig:15dpn-spec} shows the evolution of this transient into a type Ibn supernova. These SNe are likely the explosion of 
Wolf-Rayet stars which are embedded in a He-rich circumstellar medium lost by the progenitor system  \citep{2007Natur.447..829P,2007ApJ...657L.105F,2008MNRAS.389..113P}. The GMOS spectrum 
at +26 days post-peak is typical of this class with He\,{\sc i} emission lines. The 
 He\,{\sc i} $\lambda$5876\,\AA\ line has FWHM$=3000$\kms.   

To estimate the explosion epoch, 
we used a third order polynomial fit to the first six PS1 \ips-band detections to estimate a date of $57380.60\pm2.45$. This is $1.6\pm2.45$d before the detection of GW151226. The uncertainty is estimated from usage of   
different fits (order 2-4) and epochs (4 to 8). Using the first 6 epochs and a 4th order fit gives an explosion 
epoch of 57382.03, exactly  coincident with GW151226.  There is one type Ibn
which has a double peaked lightcurve \citep{2014MNRAS.443..671G} and if that were common then this method would not be accurate. 
We calculated the bolometric lightcurve after applying suitable $K-$corrections \citep{2016arXiv160401226I}. 
A comparison with the only Ibn with an early discovery and well measured rise \citep[SN~2010al;][]{2015MNRAS.449.1921P}, 
shows that the two have very different lightcurve shapes, and therefore SN~2010al is not a good template to use for dating 
the explosion. This lightcurve diversity is a feature of Ibn, likely indicating the diverse masses of the CSM and 
ejecta that power these transients \citep{2008MNRAS.389..113P,2015MNRAS.449.1921P}. 
This illustrates that PS15dpn was temporally coincident with GW151226 to within 2.45 days. 
Another unusual feature of PS15dpn is the 
detection in the radio by the VLA by \cite{GCN18873}, which is quite a luminous 6GHz  detection, 
similar to the relativistic SN~2009bb \citep{2010Natur.463..513S}. Therefore PS15dpn caught our attention because of its rarity, and also the fact that the 
remarkable pre-explosion outburst found for the 
nearest Ibn \cite[SN2006jc][]{2007Natur.447..829P}
is still quantitatively unexplained. 

Given the temporal coincidence of PS15dpn with GW151226, we estimate the probability of finding a SN Ibn
randomly in our sampled field. 
While we detected PS15dpn at $z=0.1747$, we would be sensitive to 
such an object with $M_r \simeq -19.6$ (restframe absolute magnitude) to $z=0.265$. 
Following the calculations in our first paper 
\cite[Section 6.3,][]{2016arXiv160204156S}, the cosmic rate of core-collapse SNe within $z=0.265$ implies that within 100 square degrees, there should be 3.1 CCSN explosions 
per day.  We assume an uncertainty in the explosion epoch estimate of PS15dpn of $\Delta t$  days and 
a relative rate of Ibn SNe of $\mathcal{R}_{\rm Ibn}$ (which is the fraction of core-collapse SNe that are Ibn).  Then the number of Ibn SNe within a survey area of $\mathcal{A}$ square degrees is expected to be 

\begin{equation}
N_{\rm Ibn} = 3.1\frac{\mathcal{A}}{100}\mathcal{R}_{\rm Ibn}\Delta t
\end{equation}

For $\mathcal{A}=290$, $\Delta t = 2$, $\mathcal{R}_{\rm Ibn}=0.01$, this suggests $N_{\rm Ibn}=0.18$. 
Hence the probability of a false positive (1 or more
events) when the expectation value is 0.18  is simply the Poissonian value  $p=1-\frac{\lambda^{0} e^{-\lambda}}{0!}= 1- e^{-0.18}=0.16$. In other words the probability of finding an unrelated SN Ibn exploding within 2 days of GW151226 which is unrelated and a chance coincidence is $p=0.16$. This is not convincingly low
enough to imply a causal link, but is low enough to highlight that future coincidences
should be searched for. One could argue that the appropriate value to use for $\mathcal{R}_{\rm Ibn}$ is significantly 
less than 0.01, since radio detections at this luminosity for any Ibc SN are quite rare \citep{2010Natur.463..513S}. Alternatively, 
type Ibc SNe are much more common overall, 
which would increase the value for
$\mathcal{R}_{\rm Ibn}$  significantly. One could speculate that 
a SN Ibn could potentially be  related to a GW source if it was a compact Wolf-Rayet star + BH binary, 
such as the WR+BH systems in the nearby galaxies IC10 and NGC300 \citep{2010MNRAS.403L..41C,2007ApJ...669L..21P}.  
The WR star would need to undergo core-collapse supernova, 
followed by gravitational-wave driven merger with the BH 
companion within $\sim 2$ days.  However, assuming that the star was not in contact with the black hole prior to the supernova, the merger would typically require thousands of years rather than days.  A very favourable supernova kick toward the BH could reduce the merger timescale, but would require an implausibly high kick velocity 
and/or a very low-probability kick direction. 
However the strongest argument
against a link is that the distance estimate to GW151226 
is inconsistent with the redshift of PS15dpn 
\citep{GCN18850}
The final probability density function from LIGO drops to  zero  at $z=0.1747$ ($D_L = 854$\,Mpc) as shown in the detailed companion analysis paper \citep{LV:bbh-o1}.  
While  the mechanism of \cite{2016ApJ...819L..21L} might predict a rapidly rotating massive star which   
could conceivably produce both a SN Ibn and GW emission this now seems unlikely from the calculations of \cite{2016arXiv160300511W}. Some luminous supernovae
have been explained by magnetic neutron stars 
born with millisecond periods 
\citep{2013ApJ...770..128I,2016arXiv160401226I}, and such an object would radiate gravitational 
waves if it were elliptically deformed.
However a neutron star origin is excluded as the LIGO analysis is not 
consistent with  component masses less than 4.5\msun \citep[99\% credible level;][]{gw151226}.

\section{Discussion and conclusions}

Assuming that none of the transients we 
found, including PS15dpn, are associated with GW151226 it is useful to set quantitative and meaningful upper limits on potential optical counterparts for BH-BH mergers. These also serve as a guide to our sensitivity to potential future binary neutron star (NS-NS) and neutron star -- black hole (NS-BH) merging systems that are more promising systems for producing electromagnetic counterparts particularly  redwards of  7000\AA. 

In Figure\,1 we show the 5$\sigma$ limits of every PS1 image taken during this campaign. As described in \cite{2016arXiv160204156S}, the 
 5$\sigma$ limits are calculated for each of the
51 skycells per pointing of the GPC1 camera
data products and the median per night is 
also plotted. We also plot parametrised lightcurves of three analytic 
lightcurves with timescales $t_{\rm FWHM}=4, 20, 40$\,d \citep[as defined in][]{2016arXiv160204156S}. These  
indicate detection limits of \ips=20.3, 20.8 and 20.8 respectively, or $M_i = -18, -17.5, -17.5$ at the luminosity  distance of GW151226.

Looking to the future, we plot model lightcurves 
of kilonovae from compact binary mergers (NS-NS)  of 
\cite{2015MNRAS.450.1777K} and 
\cite{2013ApJ...775...18B} 
as  illustrative examples of our survey capability
\citep[the merger models of][are also of similar luminosity]{2014ApJ...780...31T,2013ApJ...775..113T}. 
At the estimated distance of GW151226 of $D_{\rm L}\simeq440$\,Mpc, 
the predicted fluxes would be very faint (below \ips$\simeq23$). 
It is  expected that NS-NS mergers will be more
common by volume and LIGO's horizon distance for NS-NS detections is a factor of $\sim$5-10 smaller than for BH-BH mergers, depending on the BH masses
\citep{2010CQGra..27q3001A}. During the next 
science run beginning in the fall of 2016, 
LIGO is expected to be sensitive to NS-NS mergers within $D^{\rm min}_{\rm L}\lesssim75$\,Mpc 
\citep{2016arXiv160400439T} and we show in Figure\,\ref{fig:skymap} that our survey strategy would be sensitive to these. 
Ideally, our goal would be to get 0.5 mag deeper,  beginning within 24\,hrs of the GW alert. 

We further show the $i-$band lightcurve of SN~1998bw 
\citep{1998Natur.395..670G,2001ApJ...555..900P}
which is the typical energetic type Ic SN associated
with long duration gamma-ray bursts (LGRBs). 
Figure\,1 shows that if an energetic type Ic SN accompanies 
such a GRB event 
then  it would be an unambiguous, 
bright transient in our survey. We do not 
find such an object, but caution that we 
surveyed a maximum of 26.5\% of the
total LIGO probability region. 
Our results are encouraging for 
future searches for the counterparts of 
NS-NS mergers within about 100\,Mpc where
the predicted optical and near infra-red counterparts are within reach.

PS1 and ATLAS are  supported by NASA Grants NNX08AR22G, NNX12AR65G, NNX14AM74G and NNX12AR55G. Based on data from : ESO as part of PESSTO (188.D-3003, 191.D-0935), Gemini Program GN-2016A-Q-36, GN-2015B-Q-4, the UH 2.2m and the NASA/ESA Hubble Space Telescope program 14484.  We thank Gemini and HST for Directors Discretionary time.  We acknowledge EU/FP7-ERC  Grants [291222,307260,320360,615929], Weizmann-UK Making Connections Grant, 
STFC Ernest Rutherford Fellowship (KM), Sofia Kovalevskaja Award from the Alexander von Humboldt Foundation (TWC).  PS1 surveys acknowledge the PS1SC: University of Hawaii, MPIA Heidelberg, MPE Garching,  Johns Hopkins University, Durham University,  University of Edinburgh, Queen's University Belfast,  Harvard-Smithsonian CfA, LCOGT,  NCU Taiwan, STScI,   University of Maryland,  Eotvos Lorand University, Los Alamos National Laboratory, NSF Grant No. AST-1238877.

\bibliographystyle{apj}


\begin{table*}
\caption{Transients discovered by Pan-STARRS1.}
\label{tab:trans}
\begin{center}
\begin{tabular}{lllllllll}
\hline
\hline
Name & RA (J2000) & Dec (J2000)  & Disc. MJD & Disc Mag  & Spec MJD & Type & Spec $z$  & Classification source and notes \\ \hline
PS15dcq  &  03 22 55.83  &  +34 59 23.6  & 57384.29 & 19.99 & 57388.92 &  Ia       & 0.072 &  iPTF15fgy, \cite{GCN18762},\\
        &                &                &         &       &          &           &        &  \cite{GCN18807} \\ 
PS15dov$\dagger$  &  03 43 57.36  &  +39 17 43.7  & 57384.32 & 19.73 & 57386.32 &  II        & 0.016702 &              GMOS$^1$ \cite{GCN18811}\\ 
PS15dot  &  02 11 55.69  &  +13 28 17.8  & 57384.34 & 20.97 & 57386.30 &  II        & 0.149 &             GMOS \cite{GCN18811}\\  
PS15coh  &  02 15 58.45  &  +12 14 13.6  & 57384.34 & 17.72 & 57329.22 &  Ia       & 0.020 & old SN, ASASSN-15rw, \\                  &             &               &          &       &          &           &       & iPTF15fev \cite{GCN18762}, \\
&             &               &          &       &          &           &       & \cite{GCN18791}   \\ 
PS15dow  &  02 19 42.20  &  +14 09 54.7  & 57384.34 & 20.22 & 57387.21 &  Ib       & 0.05 &              GMOS \cite{GCN18811} \\ 
PS15csf  &  02 26 02.24  &  +17 03 40.4  & 57384.35 & 18.68 & 57335.18    &  II       & 0.021 &   PESSTO$^{2}$, old SN ATel\#8264  \\ 
PS15dom  &  02 34 45.62  &  +18 20 37.7  & 57384.35 & 19.01 & 57390. &  II        & 0.034 & old SN, PSN J02344555+182039, iPTF15fdv  \\
         &               &               &          &       &  &           &        &  \cite{2016ATel.8506....1P}\\
PS15don  &  02 37 11.44  &  +19 03 20.2  & 57384.35 & 20.47 & 57388.20 &  Ia        & 0.160 &             GMOS  \\ 
PS15doy  &  02 47 54.16  &  +21 46 24.0  & 57384.38 & 20.75 & 57388.23 &  Ia       & 0.190 &             GMOS  \\ 
PS15dox  &  02 40 15.05  &  +22 32 12.1  & 57384.38 & 19.23 & 57389.06 &  Ia        & 0.080 &  PESSTO, \cite{GCN18806}  \\ 
PS15dpq  &  03 09 12.74  &  +27 31 16.9  & 57384.39 & 18.84 & 57389.04 &  Ia        & 0.038 &  PESSTO, iPTF15fel, \cite{GCN18806} \\ 
         &                &                &         &       &          &           &        & \cite{GCN18791}      \\
PS15dpa  &  02 57 56.02  &  +28 53 37.1  & 57384.40 & 19.51 & 57389.03 &  Ia   & 0.079 & PESSTO \cite{GCN18806}, \\
         &               &               &          &       &           &      &       & MASTER OTJ025756.02+285337 \\
         &               &               &          &       &      &            &      &    \cite{GCN18804}  \\ 
PS15dpl  &  05 47 45.39  &  +53 36 32.4  & 57384.43 & 19.34 & 57387.40 &  Ia        & 0.03 & SN2016J, ASASSN-16ah \cite{GCN18811}\\ 
PS15dpe  &  05 44 42.66  &  +52 24 57.9  & 57384.43 & 19.44 & 57388.25 &  Ia        & 0.057 &           GMOS    \\ 
PS16ku   &  02 19 06.15  &  +10 37 45.5  & 57385.22 & 20.95 &  57401.24 & II            & 0.061 & SNIFS \\ 
PS15dpn  &  02 32 59.75  &  +18 38 07.0  & 57385.23 & 20.69 & 57387.23 &  Ibn       & 0.1747 & GMOS, \cite{GCN18811},  \\ 
&               &               &           &       &           &              &     & iPTF15fgl \cite{GCN18848}\\
&               &               &           &       &           &              &     & \cite{GCN19145},  \cite{GCN19258}\\
PS15doz  &  02 53 41.68  &  +27 29 57.8  & 57385.25 & 20.69 & ... &  ... & ...& Likely SN$^{\ast}$ , slow rise  \\ 
PS15dpc  &  03 55 46.16  &  +38 52 49.6  & 57385.27 & 20.95 & 57387.26 &  II        & 0.056 &           GMOS \cite{GCN18811}   \\ 
PS15dqc  &  05 51 13.43  &  +52 28 18.7  & 57385.29 & 21.16 &...  & ...     & ...&  Likely SN \\ 
PS15cvo  &  02 20 37.39  &  +17 02 17.9  & 57385.31 & 20.45 &...  & ...      & ...&     MASTER022037.36+170217.5, old SN \\ 
PS15dpz  &  02 40 33.01  &  +23 00 10.8  & 57385.32 & 21.15 &...  & ...     & ...& Likely SN \\ 
PS15dpb  &  03 42 23.40  &  +39 14 40.4  & 57385.36 & 20.20 & 57386.43 & II           & 0.041045 &       GMOS \cite{GCN18811} \\
PS15dpg  &  03 17 18.88  &  +32 20 06.9  & 57385.41 & 20.86 & ... & ...       & ...&   Likely SN \\ 
PS15dpx  &  06 04 35.54  &  +53 35 25.8  & 57385.56 & 20.76 &  57395.48  & ... & 0.051     &  SNIFS$^{3}$, featureless. \\ 
PS15dou  &  06 03 38.73  &  +54 41 12.1  & 57385.56 & 20.20 &  57395.51 & II         & 0.079 & SNIFS, \cite{GCN18868}      \\ 
PS15dpu  &  02 40 41.35  &  +16 49 52.0  & 57386.22 & 17.26 &  57396.88 & II         & 0.0292 &  ASASSN-15un,  \cite{GCN18868} \\ 
PS15dpt  &  02 07 34.96  &  +11 03 25.2  & 57386.22 & 20.64 & 57395.36 &  ... & ...& SNIFS, red continuum, possible foreground       \\ 
PS15dpy  &  02 28 22.75  &  +13 59 19.3  & 57386.22 & 21.31 & 57395.39 & ...           & ...& SNIFS, red continuum, possible foreground   \\ 
PS15dqa  &  02 59 41.20  &  +25 14 12.2  & 57386.24 & 20.93 & ...  &   ...          &  0.001251&  Likely LBV in NGC1156    \\ 
PS16cks & 04 22 33.25 & +43 36 53.0 &  57386.31 & 21.45 & ...  &   ... & ...  & Likely SN \\
PS15dqd  &  05 56 14.60  &  +52 51 55.2  & 57386.37 & 19.92 &...   &       ...      & ...& Likely hostless SN, 0.6$^m$ fade in 3 days\\ 
PS15dqe  &  06 05 26.88  &  +54 09 11.3  & 57386.37 & 21.51 &...   &...             & ...& Likely SN        \\ 
 PS16kv  &  02 22 53.41  &  +19 15 49.9  & 57388.22 & 21.70 &...   &    ...         & ...& Likely SN, host is SDSS J022253.48+191550.5\\ 
 PS16kx  &  02 44 42.28  &  +22 36 39.5  & 57389.32 & 21.81 &...   &    ...         & ...& Likely SN                  \\
PS16cld  & 04 51 13.33  & +48 59 21.2 &  57392.26 & 21.17  &...  &    ...         & ...&  Likely SN   \\
 PS16kw  &  02 35 50.63  &  +17 33 38.2  & 57394.22 & 21.27 & ...  &    ...         & ...& Likely SN                    \\ 
 PS16ky  &  03 22 34.61  &  +30 36 07.1  & 57397.31 & 20.89 &...   &...             & ...&  Likely SN                  \\ 
PS16bpe  &  02 38 48.30  &  +22 05 56.4  & 57397.34 & 21.56 & ...  &    ...         & ...&  Likely hostless SN        \\ 
PS16bpf  &  02 56 00.56  &  +24 48 51.8  & 57397.38 & 21.83 & ...  &        ...     & ...&  Likely SN                  \\ 
PS16bpj  &  03 29 06.15  &  +35 39 07.5  & 57397.39 & 21.82 &...   &...             & ...&   Likely SN                  \\ 
 PS16lj  &  06 23 09.10  &  +54 38 20.9  & 57397.51 & 20.66  & 57405.24 &  ...      & 0.088 & SNIFS, blue continuum, $M_i =-17.8$     \\ 
PS15bpk  &  02 37 09.56  &  +22 24 02.4  & 57402.28 & 21.33 &...   &    ...         & ...& Old SN      \\ 
PS16bpg  &  02 56 40.73  &  +27 40 12.0  & 57402.28 & 20.46 & ...  &    ...         & ...& Likely SN, rising   \\ 
PS16bps  &  06 04 34.63  &  +53 35 38.8  & 57402.39 & 21.20 & ...  &...             & ...& Likely SN                  \\ 
PS16bpu$\dagger$  &  03 43 57.13  &  +39 17 38.4  & 57402.42 & 19.34 & ...  & ...   &  0.016702 & Likely SN. Offset 2\farcs0 from position of SN2001I \\ 
PS16bpw  &  03 06 54.05  &  +28 44 23.2  & 57413.26 & 21.53 &...   &...             & ...&   Likely SN, young \\ 
PS16bqa  &  02 38 57.24  &  +18 10 40.4  & 57413.30 & 21.69 &...   &...             & ...&   Likely SN                 \\ 
PS16aeo  &  03 30 46.70  &  +36 38 23.0  & 57414.28 & 19.67 &...   &...             & ...&   Likely SN                  \\ 
PS16bpz  &  06 31 15.13  &  +54 51 52.3  & 57414.35 & 20.18 & ...  &    ...         & ...&Likely SN                    \\ 
%
%
\hline\hline
\multicolumn{5}{l}{Probably stellar variables or AGN variability}\\
\hline
PS15dpp  &  03 00 39.86  &  +28 15 25.4  & 57384.40 & 20.63 &  57395.42 & ... & ...& SDSS J030039.86+281525.4, SNIFS. QSO?\\
PS15dop  &  03 17 29.58  &  +29 34 09.2  & 57384.40 & 20.01 & ...  &...             & ...&   Likely AGN activity  \\ 
PS15dpd  &  05 09 58.63  &  +50 47 09.4  & 57384.44 & 20.34 &...   &    ...           & ...&    Likely stellar \\ 
PS15dpo  &  02 59 49.56  &  +25 10 30.4  & 57385.25 & 20.55 & ... & ...& ... &             AGN   \\ 
PS16li  &  06 18 59.16  &  +55 50 55.4  & 57397.51 & 20.18 & ...  &      ...       & ...& Likely M-dwarf flare     \\ 
PS16bpx  &  03 50 03.36  &  +37 00 52.1  & 57414.30 & 18.82 & ...  &         ...    & ...&  stellar,CSS100113-035003+370052 \\ 
%
\hline
\end{tabular}
\end{center}
$^1$ GMOS denotes classification spectra taken for this project with Gemini-N and the GMOS spectrometer with gratings either R150 or R400.\\
$^{2}$ PESSTO denotes classification spectra taken for this project with PESSTO as described in \cite{2016arXiv160204156S}.\\
$^{3}$ SNIFS denotes classification spectra taken for this project with the SNIFS instrument on the UH2.2m telescope as described in \cite{2016arXiv160204156S}.\\
$^{4}$ MJD for GW151226 is 57382.152\\
$^{\ast}$ ``Likely SN" means that the transient is not coincident with an observed point source, nor is a known 
stellar or AGN variable, and does have a candidate host galaxy nearby and a lightcurve that is consistent with being a  normal SN.\\
$\dagger$ PS15dov and PS16bpu exploded in the same galaxy UGC2836. Which also hosted SN2001I and SN2003ih.\\
\end{table*}

\end{document}